\documentclass[conference]{IEEEtran}

\usepackage{amsmath,amssymb,amsfonts}
\usepackage{stfloats}
\usepackage{cite}
\usepackage{graphicx}
\usepackage{psfrag}
\usepackage{subfigure}
\usepackage{amsmath}
\usepackage{array}
\usepackage{algorithmicx}
\usepackage{algpseudocode}
\usepackage{setspace}
\usepackage{blindtext}
\usepackage{url}
\usepackage{epstopdf}
\usepackage{verbatim}
\usepackage{color}
\usepackage{amsmath}
\usepackage{bm}
\usepackage{multirow}
\usepackage{booktabs}
\usepackage{makecell}
\usepackage{ntheorem}
\usepackage{textcomp}
\usepackage{amsmath}
\usepackage{diagbox}
\usepackage{geometry}
\def\BibTeX{{\rm B\kern-.05em{\sc i\kern-.025em b}\kern-.08em T\kern-.1667em\lower.7ex\hbox{E}\kern-.125emX}}

\newtheorem{Prob}{Problem}

\title{Self-Calibrating Indoor Localization with Crowdsourcing Fingerprints and Transfer Learning}
\author{Chenlu Xiang$^{\dagger}$, Shunqing Zhang$^{\dagger}$, Shugong Xu$^{\dagger}$, and George C. Alexandropoulos$^{\ddag}$\\
$^{\dagger}$ Shanghai Institute for Advanced Communication and Data Science, Shanghai University, China\\
$^{\ddag}$ Department of Informatics and Telecommunications, National and Kapodistrian University of Athens, Greece \\
Email: \{xcl, shunqing, shugong\}@shu.edu.cn, alexandg@di.uoa.gr}

\begin{document}
\maketitle

\begin{abstract}
Precise indoor localization is one of the key requirements for
fifth Generation (5G) and beyond, concerning various wireless
communication systems, whose applications span different vertical sectors. Although many highly accurate methods based on signal fingerprints have been lately proposed for localization, their vast majority faces the problem of degrading performance when deployed in indoor systems, where the propagation environment changes rapidly. In order to address this issue, the crowdsourcing approach has been adopted, according to which the fingerprints are frequently updated in the respective database via user reporting. However, the late crowdsourcing techniques require precise indoor floor plans and fail to provide satisfactory accuracy. In this paper, we propose a low-complexity self-calibrating indoor crowdsourcing localization system that combines historical with frequently updated fingerprints for high precision user positioning. We present a multi-kernel transfer learning approach which exploits the inner relationship between the original and updated channel measurements. Our indoor laboratory experimental results with the proposed approach and using Nexus 5 smartphones at 2.4GHz with 20MHz bandwidth have shown the feasibility of about one meter level accuracy with a reasonable fingerprint update overhead.
\end{abstract}


\section{Introduction} \label{sect:intro}
Among the technical requirements for fifth Generation (5G) wireless communications, and their late demanding provisions for 6G, belong the precise indoor localization, which can bring, as representative applications, accurate navigation experience \cite{gozick2011magnetic} in shopping malls and seamless tracking requirements in smart factories \cite{ahmed2016internet}. Different from the outdoor localization process, where the combination of the global navigation satellite system (GNSS) with the inertial navigation system (INS) \cite{hofmann2007gnss} provides satisfying accuracy, there are still diversified solutions in the indoor case, such as those based on the low-cost Bluetooth Low Energy (BLE) \cite{faragher2015location}, the increasingly popular 3GPP LTE/5G \cite{zhang2019fingerprint}, and the widely deployed WiFi technologies \cite{paul2009rssi,sen2012you,chapre2015csi,xiang2019robust,kotaru2015spotfi,vasisht2016decimeter}. Nevertheless, signal fingerprinting approaches, including large scale received signal strength indicator (RSSI) \cite{paul2009rssi}, or reference signal received power (RSRP) \cite{zhang2019fingerprint}, or small-scale channel state information (CSI) \cite{xiang2019robust,vasisht2016decimeter}, are usually recognized as the most efficient solutions for high accuracy indoor localization.

Many methods based on signal fingerprints, e.g.,  \cite{paul2009rssi,zhang2019fingerprint} and  \cite{xiang2019robust}, have achieved record-breaking localization results. Those methods target at extracting the intrinsic features of wireless signals in the training phase, which are then utilized in the online operating phase to predict the user location in conjunction with real-time measurements. The RSSI \cite{paul2009rssi} and RSRP  \cite{zhang2019fingerprint} metrics have been proven to be highly correlated with the spatial locations, and thus, deployed for improving the localization accuracy to the level of a meter in indoor and outdoor scenarios,  respectively. Furthermore, CSI measurements, which are frequently reported in \cite{sen2012you,chapre2015csi,xiang2019robust,kotaru2015spotfi,vasisht2016decimeter}, have managed to improve the localization accuracy. Those measurements have complex structure and more dimensions than traditional fingerprints. The probabilistic models between the collected CSI and the candidate locations are established through some classifiers, such as deterministic k-nearest neighbor (KNN) clustering \cite{sen2012you}, probabilistic Bayes rule algorithms \cite{chapre2015csi}, and deep learning based algorithms \cite{xiang2019robust,huang2019indoor}. The above fingerprint based solutions have been shown to be able to achieve sub-meter, even decimeter level accuracy, if CSI from multiple APs \cite{kotaru2015spotfi}, multiple frequency bands \cite{vasisht2016decimeter},  and/or multiple antennas \cite{chapre2015csi} can be fused together.

The availability of timely and accurate signal fingerprint maps is of paramount importance for the aforementioned highly accurate localization approaches. However, the collection of fingerprint maps is often considered as a labor-intensive task. Furthermore, fingerprint maps can be easily corrupted by the fluctuations in the wireless channels, due to human movement or time-varying scattering and reflections. It was shown in \cite{wang2016indoor} that those factors can gradually, over time, degrade the localization accuracy. In order to solve this problem, \cite{kalker2009fingerprint} proposed to regularly update the fingerprint database in order to maintain the positioning accuracy without deterioration. However, the associated system maintenance costs were inevitably expensive. Recently, a low cost alternative scheme named \emph{crowdsourcing} \cite{liu2019data} has been proposed to keep the fingerprint database fresh via collaborative user reporting. With the aid of floor plans, the existing crowdsourcing systems \cite{murata2019smartphone} can update the fingerprint database by matching the mobile users' zigzag routing, estimated by inertial measurement unit (IMU) results for a period of time with the pre-acquired floor plan. However, the above method suffers from the inaccurate location information of crowdsourcing users, which often results in error accumulating events, as shown in \cite{murata2019smartphone}. In addition, the localization approaches \cite{wu2014smartphones,huang2019online} only update the fingerprint database by newly collected fingerprints using conventional multi-dimensional scaling (MDS) \cite{wu2014smartphones} or marginalized particle filtering (MPF)  \cite{huang2019online} schemes, which has been shown to exhibit poor localization accuracy.

To address the aforementioned issues with the crowdsourcing approaches, we present in this paper a novel self-calibrating indoor localization system that is based on a WiFi fingerprint database. By efficiently utilizing the dynamically updated fingerprints from crowdsourcing, we introduce the maximum mean discrepancy (MMD) to describe the differences of their distributions, which is further extended to multi-kernel MMD (MK-MMD) by incorporating the conditional CSI distributions. Furthermore, a combined loss function for the proposed deep transfer learning framework is designed to balance the localization accuracy and the MK-MMD distances. The proposed scheme attains to keep the WiFi fingerprint database as well as the neural network models updated, in order to efficiently track the variations of the wireless channel for long time periods, while exhibiting reasonable implementation complexity. Our experimental results in an indoor laboratory environment have showcased that the proposed localization scheme can achieve mean localization errors as low as one meter.

The rest of this paper is organized as follows. In Section~\ref{sect:sys}, we introduce our proposed crowdsourcing indoor localization system. We then present the considered problem formulation and the implementation details of the proposed deep transfer learning approach in Section~\ref{sect:sol}. Our experimental results are discussed in Section~\ref{sect:exper}, while Section~\ref{sect:conc} concludes this paper. 

\addtolength{\topmargin}{0.2in}

\section{System Model}\label{sect:sys}
\begin{figure}
\centering
\includegraphics[width = 3.5 in]{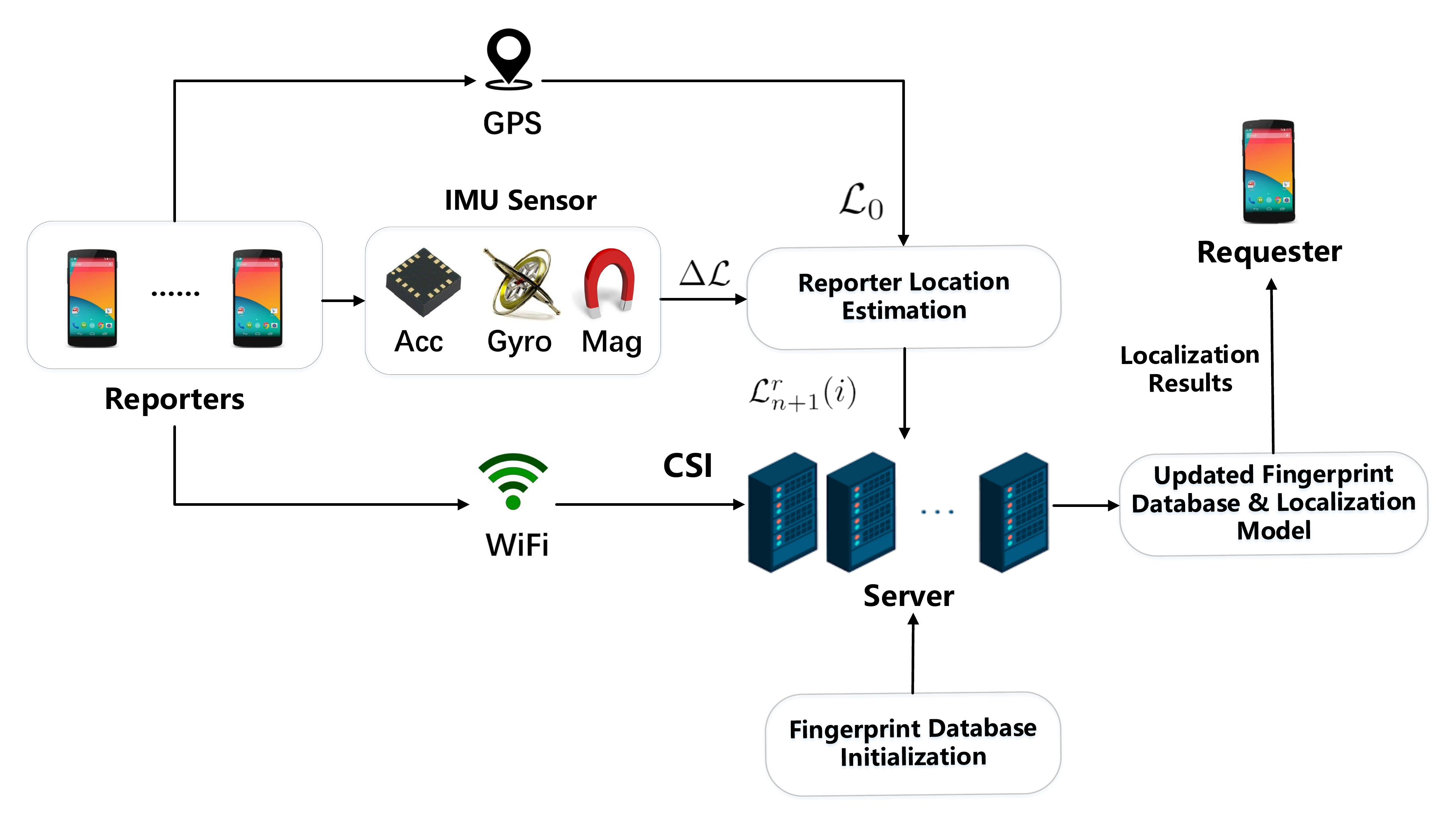}
\caption{The proposed crowdsourcing system includes one or more reporters, a server, and a requester. The system operator designs the initial fingerprint database at the server, and the reporter provides frequently updates on its fingerprints with location labels. Based on this information, the server keeps updating the relationship between the database with the fingerprints and the reporter locations. In the online phase, the requester leverages the latter updated mapping to obtain its precise localization. }
\label{fig:system}
\end{figure}

In this section, we present the proposed crowdsourcing system architecture, which is depicted in Fig.~\ref{fig:system}, and discuss the initialization of the architecture's fingerprint database as well as its updating procedure via crowdsourcing.

\subsection{Fingerprint Database Initialization}
The initial fingerprint database, whose content is denoted as $\mathcal{DB}_{0}$, stores the CSI samples from $N_{R}$ discrete grid locations, e.g., $\{\mathcal{A}_m, \forall m \in [1, \ldots, N_{R}]\}$. We make use of the notation $\mathbf{H}(\mathcal{A}_{m}, T_0) \in \mathbb{C}^{N_{sc} \times N_{s}}$ for the aggregated channel response of $N_{sc}$ subcarriers and $N_{s}$ consecutive orthogonal frequency division multiplexing (OFDM) symbols in grid area $\mathcal{A}_{m}$ at the initial time instant $T_0$. The entire fingerprint database is initialized as follows,
\begin{eqnarray}
\mathcal{DB}_{0} = \left\{\left(m, \mathbf{H}(\mathcal{A}_{m}, T_0)\right),  \forall m \in [1, \ldots, N_{R}] \right\}.
\end{eqnarray}

In practice, it is in general hard to obtain the channel responses of the entire grid area. For our fingerprint database initialization, the CSI samples at the center locations $\{\mathcal{L}_m, \forall m \in [1, \ldots, N_{R}]\}$ of each grid area (see the red points in our laboratory experimental results, as illustrated later on in Fig. ~\ref{fig:environment}), are considered as the regional CSI, $\mathbf{H}(\mathcal{A}_{m}, T_0)$, and CSI at the reference point $\mathcal{L}_{m}$, e.g., $\mathbf{H}(\mathcal{L}_{m}, T_0)$, can be obtained via a standard MMSE detection algorithm, as discussed in \cite{yang2001analysis}. 

\subsection{Fingerprint Database Update}
For each duration between the time instants $T_n$ and $T_{n+1}$, we assume $N^{r}_{n+1}$ reporters are transmitting their respective location information $\{\mathcal{L}^{r}_{n+1}(i)\}$ and collected CSI $\{\mathbf{H}(\mathcal{L}^{r}_{n+1}(i), T_{n+1})\}$, where $i \in [1, \ldots, N^{r}_{n+1}]$, to the central server. In order to obtain the accurate location $\mathcal{L}^{r}_{n+1}(i)$, we utilize the WiFi, GPS receiver and IMU sensors, as shown in Fig.~\ref{fig:system}. Let us denote by $\Omega_{i}(\mathcal{A}_{m})$ the set of reports in the area $\mathcal{A}_{m}$, e.g., $\Omega_{i}(\mathcal{A}_{m}) = \{i, \forall i \ \textrm{satisfies} \ \mathcal{L}^{r}_{n+1}(i) \in \mathcal{A}_{m}\}$. We can update the fingerprint $\mathbf{H}(\mathcal{A}_m,T_{n+1})$ by averaging the collected CSI within the area $\mathcal{A}_m$ in order to eliminate occasional measurement errors, which is given by
\begin{eqnarray*}
\mathbf{H}(\mathcal{A}_m,T_{n+1}) = \qquad \qquad \qquad \qquad \qquad \qquad \nonumber \\
\left\{
\begin{array}{cc}
\frac{1}{|\Omega_{i}(\mathcal{A}_m)|} \sum_{i \in \Omega_i(\mathcal{A}_m)} \mathbf{H}(\mathcal{L}^{r}_{n+1}(i), T_{n+1}), &  \Omega_{i}(\mathcal{A}_m) \neq \emptyset, \\
\mathbf{H}(\mathcal{A}_m,T_{n}), &  \Omega_{i}(\mathcal{A}_m) = \emptyset,
\end{array}
\right.
\end{eqnarray*}
where $|\cdot|$ denotes the cardinality of the inner set and $\emptyset$ denotes the empty set. The fingerprint database at $T_{n+1}$, i.e., $\mathcal{DB}_{n+1}$, is thus given by,
\begin{eqnarray}
\mathcal{DB}_{n+1} = \left\{\left(m, \mathbf{H}(\mathcal{A}_{m}, T_{n+1})\right),  \forall m \in [1, \ldots, N_{R}] \right\}.
\end{eqnarray}

It is well known that GPS is currently the most widely applied GNSS service in the world, and deployed on most mobile terminals. However, indoor GPS signals are usually considered too weak for indoor localization. Actually, our smartphones often receive strong GPS signals and acquire accurate positioning coordinates in some non-closed places such as positions by the windows. In this way, opportunistic GPS localization can be accessible to annotate the newly collected fingerprint. In order to make sure the received signals are strong enough to provide accurate locations, we have designed detectors to help assess the signal quality\footnote{More details about the detectors will be provided in the extended version of this paper.}. Only when the signal quality meets the requirements in several tests, we record the prior position $\mathcal{L}_{0}$. To further obtain the user traces, we apply an offline pedestrian dead reckoning (PDR) algorithm applied on the collected IMU sensor data, including accelerometer, gyroscope and magnetometer as mentioned in \cite{li2019soicp}. The whole process consists of step detection $N_{L}$, step length estimation $L_{k}$, and heading direction estimation $\alpha_k$, and the PDR displacement in a very short time period is described as,
\begin{eqnarray}
\Delta \mathcal{L} =\sum_{k=1}^{N_{L}} \alpha_k \cdot L_{k},
\end{eqnarray}
where $L_{k}$ is the $k^{th}$ step length of reporter. A particle filter \cite{huang2019online} is utilized to reduce IMU distance errors during this process. Through this approach, we can obtain the reporter's location by $\mathcal{L}^{r}_{n+1}(i) = \mathcal{L}_{0} + \Delta \mathcal{L}$ and the corresponding CSI, $\mathbf{H}(\mathcal{L}^{r}_{n+1}(i), T_{n+1})$. 

\section{Proposed Deep Transfer Learning Scheme} \label{sect:sol}
In this section, we present the considered crowdsourcing localization
problem formulation and devise a deep transfer learning scheme for the fingerprint database update. In particular, we present a MK-MMD
minimization optimization framework, based on which a novel neural network structure and loss function are designed to exploit the inner relationship between the original and updated CSI measurements.

\subsection{Problem Formulation}
We apply a general optimization framework to describe the localization problem. Denote $\mathcal{L}_n^k$ and $\hat{\mathcal{L}}_n^k$ to be the ground-true and the predicted locations of the $k^{th}$ target at $T_n$ respectively, and the corresponding {\em mean distance error} (MDE) performance over $K$ sampling positions is given by $\frac{1}{K} \sum_{k=1}^{K} \|\hat{\mathcal{L}}_n^k - \mathcal{L}_n^k\|_2$, where $\|\cdot\|_2$ represents the vector $l_2$ norm as defined in \cite{boyd2004convex}. With the above notation, we can describe the MDE minimization problem using the following optimization framework.

\begin{Prob}[MDE Minimization]
\label{prob:CBL}
\begin{eqnarray}
\underset{\{g_n(\cdot)\}}{\textrm{minimize}} && \frac{1}{N} \frac{1}{K} \sum_{n=1}^{N}\sum_{k=1}^{K} \|\hat{\mathcal{L}}_n^k - \mathcal{L}_n^k\|_2 \\
\textrm{subject to} 
&& \hat{\mathcal{L}}_{n}^{k} = \left(\mathbf{p}_n^{k}\right)^{T} \cdot \mathcal{L}_c, \\
&& \mathbf{p}_n^k = g_n \left(\mathbf{H}(\mathcal{L}_n^k),\mathcal{DB}_{n} \right), \forall n, \label{eqn:const_1} \\
&& \mathbf{p}_n^k \in [0,1]^{N_{R}}, \forall n,k,
\end{eqnarray}
where $N$ is the total number of localization time instants, as well as $\mathcal{L}_c = [\mathcal{L}_c^{1}, \ldots, \mathcal{L}_c^{m}, \ldots, \mathcal{L}_c^{N_{R}}]$ and ${\mathbf{p}_n^{k}}$ denote the central grid positions and the position likelihood distribution of the $k^{th}$ target at $T_{n}$ with respect to all the possible $\mathcal{A}_{m}$,  respectively. The function $g_n$ represents the unknown mapping relationship between the measured CSI $\mathbf{H}(\mathcal{L}_n^k)$ and ${\mathbf{p}_n^{k}}$.
\end{Prob}

In order to minimize the localization errors, it is necessary to estimate $\mathbf{p}_n^k$ by fitting the $g_n(\cdot)$ function, and further obtain a recursive relation with the new function $g_{n+1}(\cdot)$, which is represented as,
\begin{eqnarray}
\mathbf{p}_{n+1}^k = g_{n+1} \left(\mathbf{H}(\mathcal{L}_{n+1}^k),\mathcal{DB}_{n+1} \right).
\label{eqn:const_2}
\end{eqnarray}

Considering that part of fingerprints in $\mathcal{DB}_{n+1}$ are updated compared with $\mathcal{DB}_{n}$, it is unnecessary to fit the $g_{n+1}(\cdot)$ function using the entire $\mathcal{DB}_{n+1}$. In order to reduce the computational complexity of the fingerprint database update, we propose to apply transfer learning for fingerprint transfer. Due to the difference between $\textrm{Pr}(\mathbf{H}(\mathcal{A}_{m}, T_n))$ and $\textrm{Pr}(\mathbf{H}(\mathcal{A}_{m}, T_{n+1}))$, the self-calibrating localization system needs to utilize a transfer mapping function $\Phi(\cdot)$ in order to model the difference in reproducing kernel Hilbert space (RKHS) instead, where $\textrm{Pr}(\cdot)$ represents the stochastic properties of channel distributions. The corresponding MMD measure is thus given by,
\begin{eqnarray}
&\mathcal{D}_{\textrm{MMD}}(\mathcal{DB}_{n}, \mathcal{DB}_{n+1}) \nonumber \\
= & \sum_{m=1}^{N_R} \left\| \Phi\left(\mathbf{H}(\mathcal{A}_{m}, T_n))\right) -\Phi\left(\mathbf{H}(\mathcal{A}_{m}, T_{n+1}))\right)\right\|_{\mathcal{H}}^{2},\nonumber
\end{eqnarray}
where $\|\cdot\|_{\mathcal{H}}$ are and the vector norm operation in RKHS. With the above manipulation, we are required to fit the optimal mapping function $\Phi(\cdot)$ to update the function $g_{n+1}(\cdot)$, that is minimize MMD. Then we can transform the original MDE minimization problem into the following MMD minimization problem.
\begin{Prob}[MMD Minimization]
\label{prob:MMD_min}
\begin{eqnarray}
\underset{\Phi (\cdot)}{\textrm{minimize}} && \mathcal{D}_{\textrm{MMD}}(\mathcal{DB}_{n}, \mathcal{DB}_{n+1}) \nonumber \\
\textrm{subject to} && \eqref{eqn:const_1},\  \eqref{eqn:const_2}.\nonumber
\end{eqnarray}
\end{Prob}

Although the aforementioned MMD considers the distribution differences of CSI between $\textrm{Pr}(\mathbf{H}(\mathcal{A}_{m}, T_n))$ and $\textrm{Pr}(\mathbf{H}(\mathcal{A}_{m}, T_{n+1}))$, the conditional probability distributions of specific areas, e.g.,  $\textrm{Pr}(\mathbf{H}(\mathcal{A}_{m}, T_n)|\mathcal{A}_{m})$ and  $\textrm{Pr}(\mathbf{H}(\mathcal{A}_{m}, T_{n+1})|\mathcal{A}_{m})$, are ignored. Since this feature provides additional correlation information in different areas, we propose to use an improved multi-kernel solution, namely {\em MK-MMD} \cite{long2015learning}, defined as follows.
\begin{eqnarray}
& \mathcal{D}_{\textrm{MK-MMD}}(\mathcal{DB}_{n}, \mathcal{DB}_{n+1}) \nonumber \\
& = \sum_{m=1}^{N_R} \lambda \left\| \Phi\left(\mathbf{H}(\mathcal{A}_{m}, T_n))\right) -\Phi\left(\mathbf{H}(\mathcal{A}_{m}, T_{n+1}))\right)\right\|_{\mathcal{H}}^{2} \nonumber \\
& + \mu \left\| \Phi\left(\mathbf{H}(\mathcal{A}_{m}, T_n)|\mathcal{A}_{m})\right) -\Phi\left(\mathbf{H}(\mathcal{A}_{m}, T_{n+1})|\mathcal{A}_{m})\right)\right\|_{\mathcal{H}}^{2} \nonumber, \nonumber
\end{eqnarray}
where $\lambda,\mu \in [0, 1]$ denote a fine-tuning coefficient indicating the data similarity between $\mathcal{DB}_{n}$ and $\mathcal{DB}_{n+1}$, for which it holds $\lambda + \mu=1$. With the proposed MK-MMD metric, we define the MK-MMD minimization problem to better fit the optimal mapping function $\Phi(\cdot)$ and update the function $g_{n+1}(\cdot)$ as follows.

\begin{Prob}[MK-MMD Minimization]
\label{prob:MKMMD_min}
\begin{eqnarray}
\underset{\Phi(\cdot)}{\textrm{minimize}} && \mathcal{D}_{\textrm{MK-MMD}}(\mathcal{DB}_{n}, \mathcal{DB}_{n+1}) \nonumber \\
\textrm{subject to} && \lambda + \mu=1, \nonumber \\
&& \eqref{eqn:const_1},\  \eqref{eqn:const_2}.\nonumber
\end{eqnarray}
\end{Prob}

It is noted that in the conventional methods, such as joint distribution adaptation (JDA) \cite{long2013transfer}, the objective is usually to find a transformation matrix to represent the transfer mapping function $\Phi(\cdot)$. In the above MK-MMD formulation, a simply transformation matrix is insufficient, as mentioned in \cite{long2015learning}, which motivates us to apply deep neural networks for the transfer function representation, as described in the sequel.

\subsection{Neural Network}
\begin{figure}
\centering
\includegraphics[width = 3.2 in]{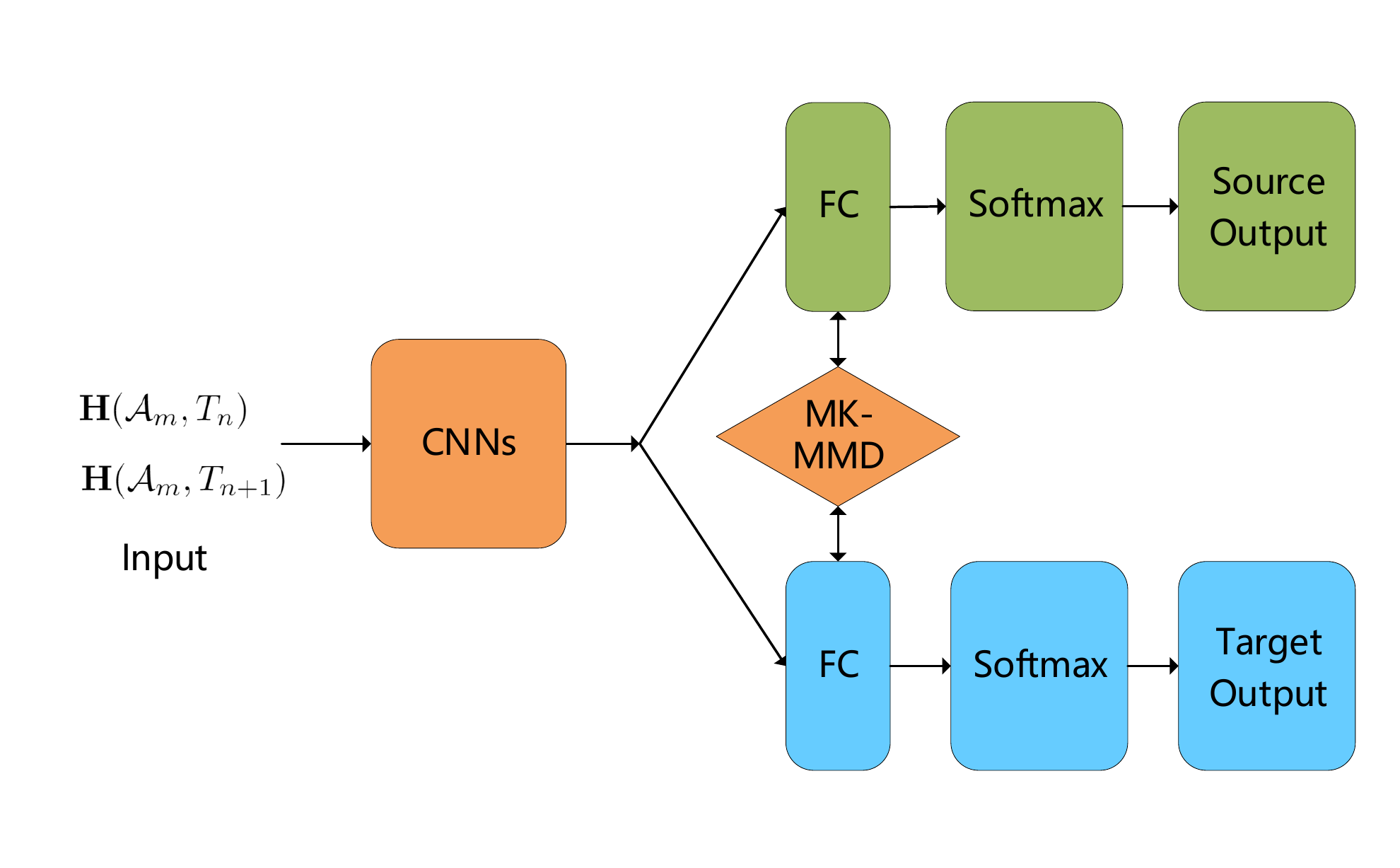}
\caption{The architecture of deep transfer learning network, which are consisted of convolution layers, FC layers and average pooling layers. The input of the network are the channel responses at the time instants $T_{n}$ and $T_{n+1}$ with respect to the grid area $\mathcal{A}_{m}$, which are $\mathbf{H}(\mathcal{A}_{m}, T_n)$ and $\mathbf{H}(\mathcal{A}_{m}, T_{n+1})$,  respectively. The output of the network is the estimated position likelihood distribution at $T_{n}$ and $T_{n+1}$, which are $\mathbf{p}_n^k$ and $\mathbf{p}_{n+1}^k$,  respectively.}
\label{fig:network}
\end{figure}

To address the mentioned Problem~\ref{prob:MKMMD_min}, we design a deep transfer learning network for fingerprint adaptation. As shown in Fig~\ref{fig:network}, we start with a deep convolutional neural network (CNN), which is a common structure to fulfill the complex signal feature extraction and dimension reduction tasks. 
Since the convolutional layers only learns generic features in the related data sets, we only impose the MK-MMD domain adaptation in the final fully connected (FC) layer. This is because the unique characteristics of different data sets begin to appear when the structure of neural networks is deep enough \cite{long2015learning}. Motivated by this fact, we design the neural network structure with five convolutional layers and one average pooling layer to obtain the feature vectors. An FC layer with softmax output \cite{jang2016categorical} is used to provide the normalized probability. The detailed configuration and parameters of the proposed neural networks are listed in Table~\ref{tab:parameter}, which is designed with reference to the commonly used transfer learning neural network. Meanwhile, in the neural network design, we also propose to use a joint loss measure, which considers both the {\em cross-entropy} measure to describe the differences between the normalized output probability $\hat{\mathbf{p}}_{n}^k$ and the ground true label vector $\mathbf{p}_{n}^k$, and the MK-MMD based loss, $\mathcal{D}_{\textrm{MK-MMD}}(\mathcal{DB}_{n}, \mathcal{DB}_{n+1})$. The parameters in the network can be adjusted by minimizing the loss function, which can be written as,
\begin{eqnarray}
\mathbb{L} = -\sum_{m=1}^{N_{R}}\mathbf{p}_{n,m}^k\log\hat{\mathbf{p}}_{n,m}^k + \mathcal{D}_{\textrm{MK-MMD}}(\mathcal{DB}_{n}, \mathcal{DB}_{n+1}),
\end{eqnarray}
where $\mathbf{p}_{n,m}^k$ and $\hat{\mathbf{p}}_{n,m}^k$ are the normalized probability for the $m^{th}$ grid area. In addition, we train the parameters of deep neural networks with Adam optimizer to minimize the above loss function in the training stage. 
\begin{table}[!ht]
\centering
\caption{An Overview of the Considered Network Configuration and Parameters.}\label{tab:parameter}
\begin{tabular}{ c c c}
\hline
\textbf{Module} & \textbf{Layers} & \textbf{Parameters} \\
\hline
\multirow{5}*{CNNs} & conv$-1$ & $112\times112\times1$ \\
& conv$-2$ & $56 \times56 \times64$  \\
& conv$-3$ & $28 \times28 \times128$  \\
& conv$-4$ & $14 \times14 \times256$  \\
& conv$-5$ & $7 \times 7 \times 512$  \\
\hline
& average pooling  & $1\times1 \times 512$  \\
& FC & $1\times1\times512$  \\
\hline
Output & Softmax & $15\times1$ \\
\hline
\end{tabular}
\end{table}

In the operating stage, users ask for their position information by reporting their real-time CSI to the server. The trained neural network with the updated parameters outputs a probability vector $\hat{\mathbf{p}}_{n+1,m}^k$, which is utilized to obtain the final estimation location $\hat{\mathcal{L}}_n^k$. The corresponding mathematical expression is given by,
\begin{eqnarray}
\hat{\mathcal{L}}_{n+1}^k = \sum_{m=1}^{N_{R}} \hat{\mathbf{p}}^{k}_{n+1,m}\cdot \mathcal{L}_{c}^{m}.
\end{eqnarray}

\section{Experiment Results} \label{sect:exper}
\begin{figure}
\centering
\includegraphics[width = 2.5 in]{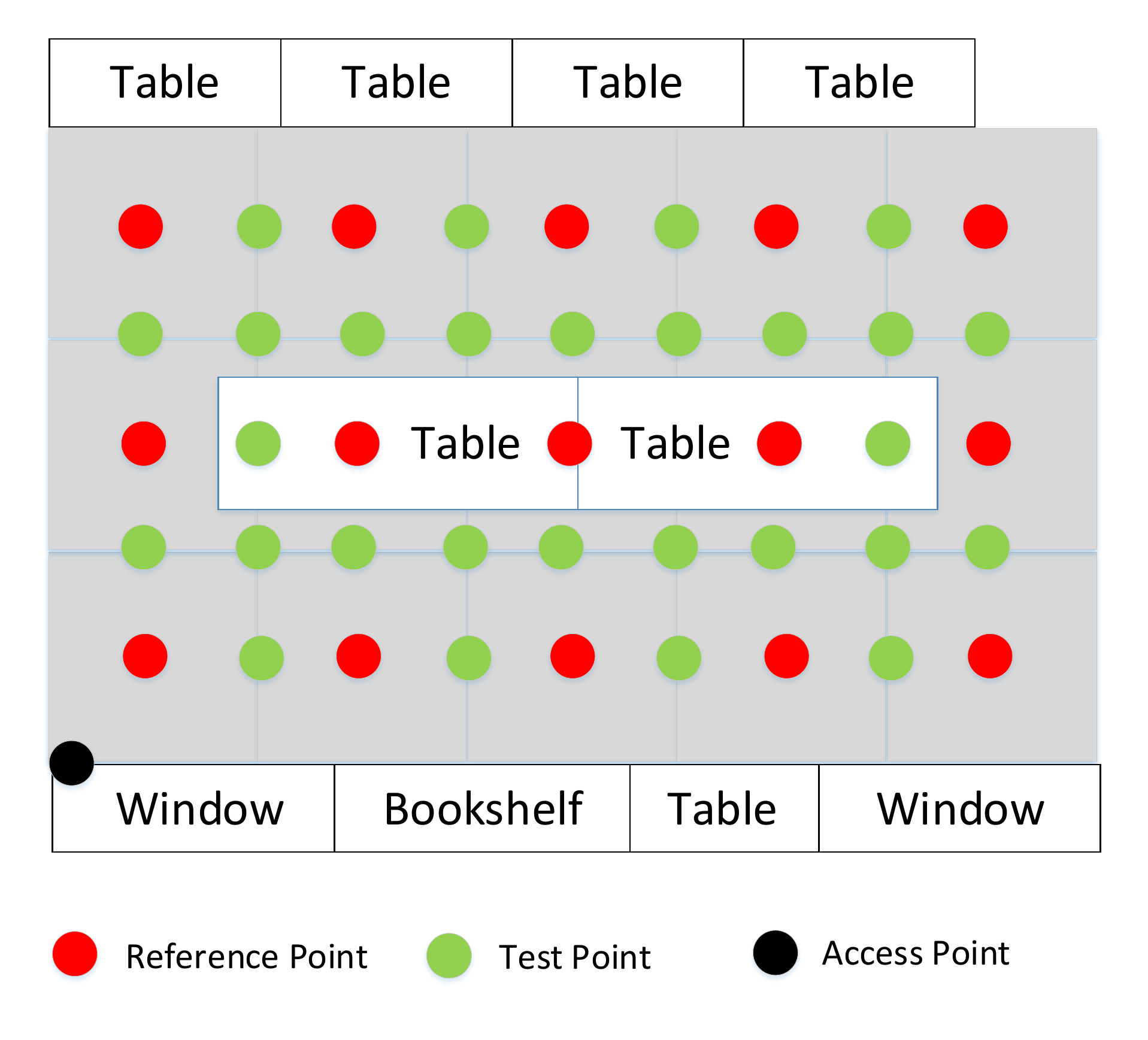}
\caption{A sketch map of experimental environment, in which the red/green/black spots represent the location of reference points of the initial fingerprint database, test points of requesters and access point, respectively. The distance between two adjacent reference points is about 1.2m. We collect 1500 CSI samples for training dataset and 750 CSI samples for test dataset.} \label{fig:environment}
\end{figure}

In this section, we provide some numerical results to show the necessity of the fingerprint database and effectiveness of the proposed deep transfer learning based approach for neural networks update. To be more specific, we compare the proposed schemes with two baselines, e.g., {\em Baseline 1}: non-updated KNN based scheme and {\em Baseline 2}: JDA based transfer scheme, in terms of effectiveness and complexity. We implement our localization system using a TP-LINK wireless router as the AP and two Nexus 5 smartphones as the reporter and requester respectively. The whole system works at 2.4 GHz with the bandwidth of 20MHz. Nexus 5 with Nexmon \cite{gringoli2019free} software installed overhears the user datagram protocol (UDP) frames transmitted by the AP and then extracts the CSI from them. We verify the proposed scheme in the laboratory environment, where the layout of testing scenarios are shown in Fig.~\ref{fig:environment}. With laboratory equipment, furniture, and people movements in the real situation, the tested wireless fading conditions cover most of the daily indoor scenarios with mixed LOS and NLOS paths.

\subsection{Localization Accuracy}
\begin{figure}
\centering
\includegraphics[width = 3.4 in]{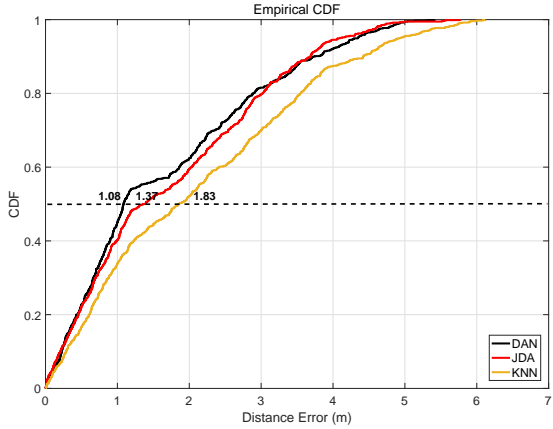}
\caption{CDF of localization distance errors for different algorithms in the test scenarios. The proposed deep transfer learning based approach is compared with two baselines to test the algorithm effectiveness.}
\label{fig:distance error}
\end{figure}

In this experiment, we investigate on the effect of fingerprint database update. We compared the proposed schemes with the above baselines by measuring the cumulative distribution function (CDF) \cite{deisenroth2020mathematics} of distance error in the test scenario. Fig.~\ref{fig:distance error} describes CDF of the localization distance error during the operating stage. The proposed regression based algorithms show superior localization accuracy over {\em Baseline 1} and {\em Baseline 2}.

By comparing JDA based approach (red solid curves) and deep transfer learning based approach (black solid curves) with non-update KNN based scheme, we can find the fingerprint transfer based schemes have a better localization performance. By comparing JDA based approach and deep transfer learning based approach, the latter one achieves the mean errors of 1.08 m for the test scenarios, which shows better localization accuracy than 1.37 m of the former one. This is due to the fact that deep transfer learning based approach is able to utilize the complex structure of neural network to better minimize MK-MDD and build the relationship between the updated fingerprint database and locations, while the transfer ability of transfer matrix in JDA based approach is very limited. 

\subsection{Fingerprint Database Update}
In this experiment, we investigate the test localization accuracy with different percentage of newly collected samples in the original dataset, considering that not all the CSI samples can be updated through crowdsourcing technology. Based on this consideration, we can figure out the time period of performing a fingerprint transfer to make sure the neural network to be updated. To this end, percentage of newly collected samples are set to be 30\%, 50\% and 70\%, respectively. 

\begin{figure}
\centering
\includegraphics[width = 3.4 in]{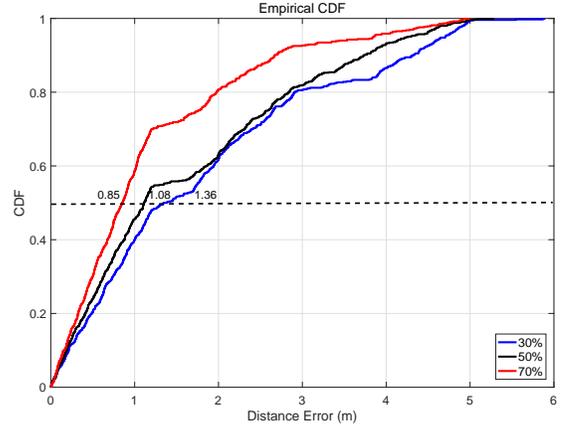}
\caption{CDF of localization errors for different percentage of newly collected samples to explore the time period of fingerprint transfer.}
\label{fig:size}
\end{figure}

Localization errors under different new CSI samples percentage are illustrated in Fig.~\ref{fig:size}, where the corresponding average localization errors are 1.36 m (blue solid curves), 1.08 m (black solid curves) and 0.85 m (red solid curves), respectively. It is worth noting that when the size is changed from 30\% to 70\%, the localization errors reduce to the half at the cost of approximately two times the data collection and labelling. Based on the above results, we believe if half of the fingerprint database can be updated by the newly collected samples, the crowdsourcing localization system can provide accuracy of about one meter.

\subsection{Computational Complexity}
Although in terms of localization errors, the proposed schemes provide a satisfactory performance, the implementation complexity is still uncertain. Therefore, in the second experiment, we show the effectiveness of proposed schemes by comparing the computational time cost with the above baseline schemes.

\begin{table}[!htbp]%
\caption{Total Running Time Comparison for Different Methods.} \label{tab:comp}
\centering
\footnotesize
\begin{tabular}{c c c c}
\toprule
\textbf{Samples Number} & \textbf{Baseline 1} & \textbf{Baseline 2 } & \textbf{Proposed Method} \\
\midrule
1 & 0.01s & 13.9s & 0.6s\\
\midrule
100 & 0.02s & 15.9s  & 4.7s \\
\midrule
1000 & 0.1s & 31.6s  & 5.61s \\
\bottomrule
\end{tabular}
\end{table}

In Table~\ref{tab:comp}, we compare the total running time of different schemes with different number of test samples on the same experimental platform. As shown in Table~\ref{tab:comp}, KNN based scheme ({\em Baseline 1}) cost the least time, regardless of the number of samples, because only the matching and  classification processes are conducted, instead of complex fingerprint transfer process. For JDA based scheme ({\em Baseline 2}), most of calculation time is used to complete the iterative process to find the suitable transfer matrix. That's why JDA algorithm cost the most running time regardless of the samples number. Thanks to the parameter storage capacity of the neural network architecture, the trained network help to quickly calculate the position information of the test signal in the online phase. Compared with two baseline methods, our proposed deep transfer learning based method provides high positioning accuracy with low computational complexity, which reduces the calculation overload of the localization system.

\section{Conclusion} \label{sect:conc}
In this paper, we presented a self-calibrating crowdsourcing localization system that is based on multi-kernel deep transfer learning for efficiently exploiting the availability of frequently updated fingerprinting signals. The proposed system simultaneously updates the fingerprint database and the corresponding localization function, enabling the utilization of both historical as well as updated fingerprint information for high precision localization. The presented indoor laboratory experimental results with the proposed system using two Nexus 5 smartphones at 2.4GHz with 20MHz bandwidth have shown that the proposed framework achieves about one meter level accuracy with relatively low computational complexity. This performance is achieved with 50\% of the fingerprint update needed in the conventional transfer matrix based method. 

\section*{Acknowledgement}
This work was supported in part by the National Natural Science Foundation of China (NSFC) Grants under No. 61701293 and No. 61871262, the National Science and Technology Major Project (Grant No. 2018ZX03001009), the National Key Research and Development Program of China (Grant No. 2017YFE0121400), the Huawei Innovation Research Program (HIRP), and research funds from Shanghai Institute for Advanced Communication and Data Science (SICS).

\bibliographystyle{IEEEtran}
\bibliography{IEEEabrv,wifi}

\begin{thebibliography}{10}
\providecommand{\url}[1]{#1}
\csname url@samestyle\endcsname
\providecommand{\newblock}{\relax}
\providecommand{\bibinfo}[2]{#2}
\providecommand{\BIBentrySTDinterwordspacing}{\spaceskip=0pt\relax}
\providecommand{\BIBentryALTinterwordstretchfactor}{4}
\providecommand{\BIBentryALTinterwordspacing}{\spaceskip=\fontdimen2\font plus
\BIBentryALTinterwordstretchfactor\fontdimen3\font minus
  \fontdimen4\font\relax}
\providecommand{\BIBforeignlanguage}[2]{{%
\expandafter\ifx\csname l@#1\endcsname\relax
\typeout{** WARNING: IEEEtran.bst: No hyphenation pattern has been}%
\typeout{** loaded for the language `#1'. Using the pattern for}%
\typeout{** the default language instead.}%
\else
\language=\csname l@#1\endcsname
\fi
#2}}
\providecommand{\BIBdecl}{\relax}
\BIBdecl

\bibitem{gozick2011magnetic}
B.~Gozick, K.~P. Subbu, R.~Dantu, and T.~Maeshiro, ``{Magnetic maps for indoor
  navigation},'' \emph{IEEE Trans. Instrum. Meas.}, vol.~60, no.~12, pp.
  3883--3891, Dec. 2011.

\bibitem{ahmed2016internet}
E.~Ahmed, I.~Yaqoob, A.~Gani, M.~Imran, and M.~Guizani,
  ``{Internet-of-things-based smart environments: state of the art, taxonomy,
  and open research challenges},'' \emph{IEEE Wireless Commun.}, vol.~23,
  no.~5, pp. 10--16, Oct. 2016.

\bibitem{hofmann2007gnss}
B.~Hofmann-Wellenhof, H.~Lichtenegger, and E.~Wasle, \emph{{GNSS--Global
  navigation satellite systems: GPS, GLONASS, Galileo, and more}}.\hskip 1em
  plus 0.5em minus 0.4em\relax Springer Science \& Business Media, 2007.

\bibitem{faragher2015location}
R.~Faragher and R.~Harle, ``{Location fingerprinting with bluetooth low energy
  beacons},'' \emph{IEEE J. Sel. Areas in Commun.}, vol.~33, no.~11, pp.
  2418--2428, Nov. 2015.

\bibitem{zhang2019fingerprint}
H.~Zhang, Z.~Zhang, S.~Zhang, S.~Xu, and S.~Cao, ``{Fingerprint-based
  localization using commercial LTE signals: A field-trial study},'' in
  \emph{IEEE Veh. Technol. Conf.}\hskip 1em plus 0.5em minus 0.4em\relax IEEE,
  2019, pp. 1--5.

\bibitem{paul2009rssi}
A.~S. Paul and E.~A. Wan, ``{RSSI-based indoor localization and tracking using
  Sigma-point Kalman smoothers},'' \emph{IEEE J. of Sel. Topics Signal
  Process.}, vol.~3, no.~5, pp. 860--873, Oct. 2009.

\bibitem{sen2012you}
S.~Sen, B.~Radunovic, R.~R. Choudhury, and T.~Minka, ``You are facing the
  {M}ona {L}isa: spot localization using phy layer information,'' in \emph{ACM
  Proc. MobiSys'12}, 2012, pp. 183--196.

\bibitem{chapre2015csi}
Y.~Chapre, A.~Ignjatovic, A.~Seneviratne, and S.~Jha, ``{CSI-MIMO: An efficient
  Wi-Fi fingerprinting using channel state information with MIMO},''
  \emph{Perv. Mobile Comput.}, vol.~23, pp. 89--103, 2015.

\bibitem{xiang2019robust}
C.~Xiang, S.~Zhang, S.~Xu, X.~Chen, S.~Cao, G.~C. Alexandropoulos, and V.~K.
  Lau, ``{Robust sub-meter level indoor localization with a single WiFi access
  point—regression versus classification},'' \emph{IEEE Access}, vol.~7, pp.
  146\,309--146\,321, 2019.

\bibitem{kotaru2015spotfi}
M.~Kotaru, K.~Joshi, D.~Bharadia, and S.~Katti, ``{SpotFi: Decimeter level
  localization using WiFi},'' in \emph{Proc. ACM Conf. Special Interest Group
  Data Commun.}\hskip 1em plus 0.5em minus 0.4em\relax ACM, 2015, pp. 269--282.

\bibitem{vasisht2016decimeter}
D.~Vasisht, S.~Kumar, and D.~Katabi, ``{Decimeter-level localization with a
  single WiFi access point},'' in \emph{Proc. Usenix Conf. Netw. Syst. Des.
  Implementation}.\hskip 1em plus 0.5em minus 0.4em\relax ACM, 2016, pp.
  165--178.

\bibitem{huang2019indoor}
C.~Huang, G.~C. Alexandropoulos, C.~Yuen, and M.~Debbah, ``Indoor signal
  focusing with deep learning designed reconfigurable intelligent surfaces,''
  in \emph{Proc. Int. Workshop Signal Process. Adv. Wireless Commun.}\hskip 1em
  plus 0.5em minus 0.4em\relax IEEE, 2019, pp. 1--5.

\bibitem{wang2016indoor}
B.~Wang, Q.~Chen, L.~T. Yang, and H.-C. Chao, ``{Indoor smartphone localization
  via fingerprint crowdsourcing: challenges and approaches},'' \emph{IEEE
  Wireless Commun.}, vol.~23, no.~3, pp. 82--89, Jun. 2016.

\bibitem{kalker2009fingerprint}
A.~A. C.~M. Kalker and J.~A. Haitsma, ``{Fingerprint database updating method,
  client and server},'' Apr. 2009, {U.S. Patent 7,523,312}.

\bibitem{liu2019data}
Y.~Liu, L.~Kong, and G.~Chen, ``{Data-oriented mobile crowdsensing: A
  comprehensive survey},'' \emph{IEEE Commun. Surv. Tut.}, vol.~21, no.~3, pp.
  2849--2885, 3rd Quarter 2019.

\bibitem{murata2019smartphone}
M.~Murata, D.~Ahmetovic, D.~Sato, H.~Takagi, K.~M. Kitani, and C.~Asakawa,
  ``{Smartphone-based localization for blind navigation in building-scale
  indoor environments},'' \emph{Perv. Mobile Comput.}, vol.~57, pp. 14--32,
  2019.

\bibitem{wu2014smartphones}
C.~Wu, Z.~Yang, and Y.~Liu, ``{Smartphones based crowdsourcing for indoor
  localization},'' \emph{IEEE Trans. Mobile Comput.}, vol.~14, no.~2, pp.
  444--457, Feb. 2014.

\bibitem{huang2019online}
B.~Huang, Z.~Xu, B.~Jia, and G.~Mao, ``{An online radio map update scheme for
  WiFi fingerprint-based localization},'' \emph{IEEE Int. Things J.}, vol.~6,
  no.~4, pp. 6909--6918, Aug. 2019.

\bibitem{yang2001analysis}
B.~Yang, Z.~Cao, and K.~B. Letaief, ``{Analysis of low-complexity windowed
  DFT-based MMSE channel estimator for OFDM systems},'' \emph{IEEE Trans.
  Commun.}, vol.~49, no.~11, pp. 1977--1987, 2001.

\bibitem{li2019soicp}
Z.~Li, X.~Zhao, F.~Hu, Z.~Zhao, J.~L.~C. Villacr{\'e}s, and T.~Braun, ``{SoiCP:
  A seamless outdoor--Indoor crowdsensing positioning system},'' \emph{IEEE
  Int. Things J.}, vol.~6, no.~5, pp. 8626--8644, 2019.

\bibitem{boyd2004convex}
S.~Boyd, S.~P. Boyd, and L.~Vandenberghe, \emph{{Convex optimization}}.\hskip
  1em plus 0.5em minus 0.4em\relax Cambridge University Press, 2004.

\bibitem{long2015learning}
M.~Long, Y.~Cao, J.~Wang, and M.~I. Jordan, ``{Learning transferable features
  with deep adaptation networks},'' in \emph{Proc. Int. Conf. Mach.
  Learn.}\hskip 1em plus 0.5em minus 0.4em\relax ACM, 2015.

\bibitem{long2013transfer}
M.~Long, J.~Wang, G.~Ding, J.~Sun, and P.~S. Yu, ``{Transfer feature learning
  with joint distribution adaptation},'' in \emph{Proc. Int. Conf. Comput.
  Vision}.\hskip 1em plus 0.5em minus 0.4em\relax IEEE, 2013, pp. 2200--2207.

\bibitem{jang2016categorical}
E.~Jang, S.~Gu, and B.~Poole, ``{Categorical reparameterization with
  Gumbel-softmax},'' \emph{arXiv preprint arXiv:1611.01144}, 2016.

\bibitem{gringoli2019free}
F.~Gringoli, M.~Schulz, J.~Link, and M.~Hollick, ``{Free your CSI: A channel
  state information extraction platform for modern Wi-Fi chipsets},'' in
  \emph{Proc. Int. Workshop Wireless Net. Testbeds, Exper. Eval. Character.},
  2019, pp. 21--28.

\bibitem{deisenroth2020mathematics}
M.~P. Deisenroth, A.~A. Faisal, and C.~S. Ong, \emph{{Mathematics for machine
  learning}}.\hskip 1em plus 0.5em minus 0.4em\relax Cambridge University
  Press, 2020.

\end{thebibliography}
\end{document}